\documentclass{jpsj3}
\usepackage{txfonts}
\usepackage{bm}
\usepackage{enumerate}
\usepackage{amssymb,graphicx,amsmath}

\newtheorem{definition}{Definition}

\newtheorem{theorem}{Theorem}

\title{Necessary and Sufficient Condition for the Existence of Zero-Determinant Strategies in Repeated Games}

\author{Masahiko Ueda$^1$\thanks{m.ueda@yamaguchi-u.ac.jp}}
\inst{$^1$Graduate School of Sciences and Technology for Innovation, Yamaguchi University, Yamaguchi 753-8511, Japan} 

\abst{Zero-determinant strategies are a class of memory-one strategies in repeated games which unilaterally enforce linear relationships between payoffs.
It has long been unclear for what stage games zero-determinant strategies exist.
We provide a necessary and sufficient condition for the existence of zero-determinant strategies.
This condition can be interpreted as the existence of two different actions which unilaterally adjust the total value of a linear combination of payoffs.
A relation between the class of stage games where zero-determinant strategies exist and other class of stage games is also provided.
}

\begin{document}
\maketitle

\section{Introduction}
\label{sec:intro}
Zero-determinant (ZD) strategies are a class of memory-one strategies (strategies which recall only one previous period) in repeated games which unilaterally enforce linear relationships between payoffs of players.
ZD strategies were first discovered by two physicists, Press and Dyson, in the repeated prisoner's dilemma games \cite{PreDys2012}.
ZD strategies contain several counterintuitive examples, such as the equalizer strategy, which unilaterally sets the payoff of the opponent, and the extortionate strategy, which always obtains payoff greater than or equal to that of the opponent.
ZD strategies also contain the generous ZD strategy, which achieves a cooperative Nash equilibrium \cite{Aki2016}.
After their discovery, many extensions have been done mainly in two directions.
The first direction is extension of the range of application of ZD strategies.
Concretely, ZD strategies were extended to multi-player multi-action games \cite{HWTN2014,PHRT2015,McAHau2016,HDND2016}, games with a discount factor \cite{HTS2015,McAHau2016,IchMas2018}, games with imperfect monitoring \cite{HRZ2015,MamIch2019,UedTan2020,MamIch2020}, and games with asynchronous update \cite{McAHau2017}.
The second direction is extension of the ability of payoff control.
The concept of ZD strategies was extended so as to control moments of payoffs \cite{Ued2021}, time correlation functions of payoffs \cite{Ued2021b}, and conditional expectations of payoffs \cite{Uedsub}.
A mathematical framework of ZD strategies has been used to classify memory-one strategies into such as partner strategies and rival strategies, in social dilemma situation \cite{Aki2016,StePlo2013,HTS2015}.
Furthermore, the relation between unbeatable imitation \cite{DOS2012b,DOS2014} and ZD strategies has gradually been clarified in two-player symmetric games \cite{Ued2022}.

Although ZD strategies have been found in several stage games, such as the prisoner's dilemma game \cite{PreDys2012}, the public goods game \cite{HWTN2014,PHRT2015}, the continuous donation game \cite{McAHau2016}, a two-player two-action asymmetric game \cite{TahGho2020}, and two-player symmetric potential games \cite{Ued2022}, a condition for the existence of ZD strategies has not been clear.
For example, it has been known that ZD strategies do not exist in the rock-paper-scissors game \cite{UedTan2020}.
It has been believed that the existence of ZD strategies is highly dependent on the structure of the stage game.

In this paper, we provide a necessary and sufficient condition for the existence of ZD strategies.
This condition implies that the stage game must be easy to handle in some sense for players who want to use ZD strategies for the existence of ZD strategies.
From another perspective, we can introduce a class of stage games in which ZD strategies exist.
Such classification of stage games may be useful similarly as symmetric games \cite{Pla2017}, potential games \cite{MonSha1996}, and generalized rock-paper-scissors games \cite{DOS2012b}.
We provide a relation between the class of stage games where ZD strategies exist and other class of games, for the case of two-player symmetric games.

This paper is organized as follows.
In section \ref{sec:preliminaries}, we introduce repeated games and ZD strategies.
In section \ref{sec:result}, we provide our main theorem about the necessary and sufficient condition for the existence of ZD strategies.
A relation between the class of stage games where ZD strategies exist and other class of stage games is also provided in the section.
Section \ref{sec:conclusion} is devoted to concluding remarks.

\section{Preliminaries}
\label{sec:preliminaries}
We consider a repeated game \cite{MaiSam2006}.
The set of players is described as $\mathcal{N}:=\left\{ 1, \cdots, N \right\}$, where $N>1$ is the number of players.
The action of player $j\in \mathcal{N}$ in the stage game is written as $\sigma_j \in A_j := \left\{ 1, \cdots, M_j \right\}$, where $M_j$ is a natural number describing the number of action of player $j$.
We collectively write $\mathcal{A}:=\prod_{j=1}^N A_j$ and $\bm{\sigma}:=\left( \sigma_1, \cdots,  \sigma_N \right)\in \mathcal{A}$.
We call $\bm{\sigma}$ an action profile.
The payoff of player $j$ when the action profile is $\bm{\sigma}$ is described as $s_j\left( \bm{\sigma} \right)$.
Therefore, the stage game is $G:= \left( \mathcal{N}, \left\{ A_j \right\}_{j\in \mathcal{N}}, \left\{ s_j \right\}_{j\in \mathcal{N}} \right)$.
We write a probability $M$-simplex by $\Delta_M$.
We also introduce the notation $\sigma_{-j} := \left( \sigma_1, \cdots, \sigma_{j-1}, \sigma_{j+1}, \cdots, \sigma_N \right)\in \prod_{k\neq j} A_k$.

We repeat the stage game $G$ infinitely.
We write an action of player $j$ at round $t\geq 1$ as $\sigma_j^{(t)}$.
The behavior strategy of player $j$ is described as $\mathcal{T}_j := \left\{ T^{(t)}_j \right\}_{t=1}^\infty$, where $T^{(t)}_j: \mathcal{A}^{t-1} \to \Delta_{M_j}$ is the conditional probability at $t$-th round.
We write the expectation of the quantity $B$ with respect to strategies of all players by $\mathbb{E}[B]$.
We introduce a discounting factor $\delta$ satisfying $0\leq \delta \leq 1$ in order to discount future payoffs.
The payoff of player $j$ in the infinitely repeated game is defined by
\begin{eqnarray}
 \mathcal{S}_j &:=& \left\{
  \begin{array}{ll}
    (1-\delta) \sum_{t=1}^\infty \delta^{t-1} \mathbb{E} \left[ s_j\left( \bm{\sigma}^{(t)} \right) \right] & \left( 0\leq \delta < 1 \right) \\
    \lim_{T\rightarrow \infty} \frac{1}{T} \sum_{t=1}^T \mathbb{E} \left[ s_j\left( \bm{\sigma}^{(t)} \right) \right] & \left( \delta=1 \right).
  \end{array}
  \right.
\end{eqnarray}
In this paper, we consider only the case $\delta=1$ \cite{PreDys2012}.

A time-independent memory-one strategy of player $j$ is defined as a strategy such that $T^{(t)}_j=T_j$ for $\forall t\geq 2$ and $\sigma_j^{(t)}$ is determined only by $\bm{\sigma}^{(t-1)}$.
For time-independent memory-one strategies $T_j$ of player $j$, we introduce the Press-Dyson vectors \cite{Aki2016,UedTan2020}
\begin{eqnarray}
 \hat{T}_j\left( \sigma_j | \bm{\sigma}^{\prime} \right) &:=& T_j\left( \sigma_j | \bm{\sigma}^{\prime} \right) -  \delta_{\sigma_j, \sigma^{\prime}_j} \quad \left( \forall \sigma_j, \forall \bm{\sigma}^{\prime} \right),
 \label{eq:PD}
\end{eqnarray}
where $\delta_{\sigma, \sigma^\prime}$ is the Kronecker delta.
The second term in the right-hand side of Eq. (\ref{eq:PD}) can be regarded as a memory-one strategy (called ``Repeat'') which repeats his/her own previous action, and therefore the Press-Dyson vectors are interpreted as the difference between his/her own strategy and ``Repeat''.
It should be noted that, due to the properties of the conditional probability $T_j$, the Press-Dyson vectors satisfy several relations.
First, it satisfies
\begin{eqnarray}
 \sum_{\sigma_j} \hat{T}_j \left( \sigma_j | \bm{\sigma}^\prime \right) &=& 0 \quad \left( \forall \bm{\sigma}^\prime \right)
 \label{eq:PD_normalized}
\end{eqnarray}
due to the normalization condition of $T_j$.
Second, it satisfies
\begin{eqnarray}
 \hat{T}_j \left( \sigma_j | \bm{\sigma}^\prime \right) && \left\{
  \begin{array}{ll}
    \leq 0, & \left(\sigma_j = \sigma^\prime_j \right) \\
    \geq 0, & \left(\sigma_j \neq \sigma^\prime_j \right)
  \end{array}
  \right.
  \label{eq:property_strategy}
\end{eqnarray}
for all $\sigma_j$ and all $\bm{\sigma}^\prime$.
Third, it satisfies
\begin{eqnarray}
 \left| \hat{T}_j \left( \sigma_j | \bm{\sigma}^\prime \right) \right| &\leq& 1 \quad \left( \forall \sigma_j, \forall \bm{\sigma}^{\prime} \right).
 \label{eq:condition_strategy}
\end{eqnarray}
The last two comes from the fact that $T_j$ takes value in $[0,1]$.

For simplicity, we introduce the notation $s_0\left( \bm{\sigma} \right)=1$ $(\forall \bm{\sigma})$.
By using the Press-Dyson vectors, we define the zero-determinant strategies.
\begin{definition}[\cite{PreDys2012,McAHau2016}]
\label{def:ZDS}
A time-independent memory-one strategy of player $j$ is a \emph{zero-determinant (ZD) strategy} when its Press-Dyson vectors can be written in the form
\begin{eqnarray}
 \sum_{\sigma_j} c_{\sigma_j} \hat{T}_j\left( \sigma_j | \bm{\sigma}^{\prime} \right) &=& \sum_{k=0}^N \alpha_{k} s_{k} \left( \bm{\sigma}^{\prime} \right) \quad \left( \forall \bm{\sigma}^{\prime} \right)
 \label{eq:ZDS}
\end{eqnarray}
with some nontrivial coefficients $\left\{ \alpha_{k} \right\}$ and $\left\{ c_{\sigma_j} \right\}$ (that is, not $\alpha_0=\alpha_1=\cdots=\alpha_N=0$, and not $c_1=\cdots=c_{M_j}=\mathrm{const.}$) and Eq. (\ref{eq:ZDS}) is not zero for some $\bm{\sigma}^{\prime}$.
\end{definition}
In other words, in ZD strategies, a linear combination of the Press-Dyson vectors is described as a linear combination of payoff vectors and a vector of all ones.
It has been known that a ZD strategy (\ref{eq:ZDS}) unilaterally enforces a linear relation between expected payoffs \cite{PreDys2012,Aki2016,Ued2022}:
\begin{eqnarray}
 0 &=& \sum_{k=0}^N \alpha_{k} \left\langle s_{k} \right\rangle^{*},
 \label{eq:linear}
\end{eqnarray}
where $\left\langle \cdots \right\rangle^{*}$ is the expectation with respect to the limit-of-means distribution
\begin{eqnarray}
 P^* \left( \bm{\sigma} \right) &:=& \lim_{T\rightarrow \infty} \frac{1}{T} \sum_{t=1}^T P_{t} \left( \bm{\sigma} \right) \quad (\forall \bm{\sigma}),
\end{eqnarray}
and 
\begin{eqnarray}
 P_t \left( \bm{\sigma}^{(t)} \right) &:=& \sum_{\bm{\sigma}^{(t-1)}} \cdots \sum_{\bm{\sigma}^{(1)}} P\left( \bm{\sigma}^{(t)}, \cdots, \bm{\sigma}^{(1)} \right)
\end{eqnarray}
is the marginal distribution obtained from the joint distribution of action profiles.
Because $\mathcal{S}_k=\left\langle s_{k} \right\rangle^{*}$ $(\forall k)$, the linear relation (\ref{eq:linear}) can be interpreted as a linear relation between payoffs in the repeated game.

\section{Results}
\label{sec:result}
\subsection{Necessary and sufficient condition for the existence of ZD strategies}
\label{subsec:necsuf}
Although ZD strategies have been found in several stage games, such as the prisoner's dilemma game \cite{PreDys2012}, the public goods game \cite{HWTN2014,PHRT2015}, the continuous donation game \cite{McAHau2016}, a two-player two-action asymmetric game \cite{TahGho2020}, and two-player symmetric potential games \cite{Ued2022}, the condition of the existence of ZD strategies has not been clear.
In this section, we provide a necessary and sufficient condition for the existence of ZD strategies.

\begin{theorem}
\label{thr:existence}
A ZD strategy of player $j$ exists if and only if there exist some nontrivial coefficients $\{ \alpha_k \}_{k=0}^N$ and two different actions $\overline{\sigma}_j, \underline{\sigma}_j \in A_j$ of player $j$ such that
\begin{eqnarray}
 \sum_{k=0}^N \alpha_{k} s_{k} \left( \overline{\sigma}_j, \sigma_{-j} \right) &\leq& 0 \quad \left( \forall \sigma_{-j} \right) \nonumber \\
 \sum_{k=0}^N \alpha_{k} s_{k} \left( \underline{\sigma}_j, \sigma_{-j} \right) &\geq& 0 \quad \left( \forall \sigma_{-j} \right),
 \label{eq:condition_exsitence}
\end{eqnarray}
and $\sum_{k=0}^N \alpha_{k} s_{k}$ is not identically zero, for the stage game $G$.
\end{theorem}

\textit{Proof.}
(Necessity) 
If a ZD strategy of player $j$ exists, then the Press-Dyson vectors satisfy
\begin{eqnarray}
 \sum_{\sigma_j} c_{\sigma_j} \hat{T}_j\left( \sigma_j | \bm{\sigma}^{\prime} \right) &=& \sum_{k=0}^N \alpha_{k} s_{k} \left( \bm{\sigma}^{\prime} \right) \quad \left( \forall \bm{\sigma}^{\prime} \right)
 \label{eq:ZDS_thm}
\end{eqnarray}
with some nontrivial coefficients $\left\{ \alpha_{k} \right\}$ and $\left\{ c_{\sigma_j} \right\}$ and Eq. (\ref{eq:ZDS_thm}) is not identically zero.
Below we write $B\left( \bm{\sigma} \right):=\sum_{k=0}^N \alpha_{k} s_{k}\left( \bm{\sigma} \right)$ $(\forall \bm{\sigma})$.
By using Eq. (\ref{eq:PD_normalized}), this can be written as
\begin{eqnarray}
 B\left( \bm{\sigma}^{\prime} \right) &=& \sum_{\sigma_j} \left( c_{\sigma_j} - c_\mathrm{max} \right) \hat{T}_j\left( \sigma_j | \bm{\sigma}^{\prime} \right)
\end{eqnarray}
and 
\begin{eqnarray}
 B\left( \bm{\sigma}^{\prime} \right) &=& \sum_{\sigma_j} \left( c_{\sigma_j} - c_\mathrm{min} \right) \hat{T}_j\left( \sigma_j | \bm{\sigma}^{\prime} \right),
\end{eqnarray}
where we have defined
\begin{eqnarray}
 c_\mathrm{max} &:=& \max_{\sigma_j} c_{\sigma_j} \\
 c_\mathrm{min} &:=& \min_{\sigma_j} c_{\sigma_j}.
\end{eqnarray}
We also introduce
\begin{eqnarray}
 \sigma_\mathrm{max} &:=& \arg \max_{\sigma_j} c_{\sigma_j} \\
 \sigma_\mathrm{min} &:=& \arg \min_{\sigma_j} c_{\sigma_j},
\end{eqnarray}
where ties may be broken arbitrarily.
It should also be noted that $\sigma_\mathrm{max} \neq \sigma_\mathrm{min}$, because the left-hand-side of Eq. (\ref{eq:ZDS_thm}) becomes $0$ if $\sigma_\mathrm{max} = \sigma_\mathrm{min}$ and therefore $c_\mathrm{max}=c_\mathrm{min}$, which contradicts with the definition of ZD strategies.
Then, by using the property (\ref{eq:property_strategy}), we obtain
\begin{eqnarray}
 B\left( \sigma_\mathrm{max}, \sigma_{-j}^\prime \right) &=& \sum_{\sigma_j} \left( c_{\sigma_j} - c_\mathrm{max} \right) \hat{T}_j\left( \sigma_j | \sigma_\mathrm{max}, \sigma_{-j}^\prime \right) \nonumber \\
 &\leq& 0 \quad \left( \forall \sigma_{-j}^\prime \right)
\end{eqnarray}
and 
\begin{eqnarray}
 B\left( \sigma_\mathrm{min}, \sigma_{-j}^\prime \right) &=& \sum_{\sigma_j} \left( c_{\sigma_j} - c_\mathrm{min} \right) \hat{T}_j\left( \sigma_j | \sigma_\mathrm{min}, \sigma_{-j}^\prime \right) \nonumber \\
 &\geq& 0 \quad \left( \forall \sigma_{-j}^\prime \right).
\end{eqnarray}
Therefore, the ZD strategy satisfies the condition (\ref{eq:condition_exsitence}) with $\overline{\sigma}_j=\sigma_\mathrm{max}$ and $\underline{\sigma}_j=\sigma_\mathrm{min}$.

(Sufficiency) 
If there exist some nontrivial coefficients $\{ \alpha_k \}_{k=0}^N$ and two different actions $\overline{\sigma}_j$ and $\underline{\sigma}_j$ of player $j$ satisfying the condition (\ref{eq:condition_exsitence}), we can construct a ZD strategy as follows.
We first introduce $M:= \prod_{k=1}^N M_k$ and a vector notation of a $M$-component quantity $D(\bm{\sigma}) \in \mathbb{R}$ by $\bm{D}:= \left( D(\bm{\sigma}) \right)_{\bm{\sigma}\in \mathcal{A}} \in \mathbb{R}^M$.
We also introduce vectors obtained from $\bm{D}$
\begin{eqnarray}
 \left[ \bm{D} \right]_{\sigma_j, d} &:=& \left( D(\bm{\sigma}^\prime) \mathbb{I}(dD(\bm{\sigma}^\prime)> 0) \mathbb{I}(\sigma_j^\prime=\sigma_j) \right)_{\bm{\sigma}^\prime \in \mathcal{A}} \quad \left( \sigma_j\in A_j, d\in \{ +,- \} \right), \nonumber \\
 &&
\end{eqnarray}
where $\mathbb{I}(\cdots)$ is an indicator function which returns $1$ if $\cdots$ holds, and $0$ otherwise. 
By the definition, any $M$-component vectors $\bm{D}$ can be decomposed into linearly independent vectors
\begin{eqnarray}
 \bm{D} &=& \sum_{\sigma_j} \sum_{d=+,-} \left[ \bm{D} \right]_{\sigma_j, d}.
\end{eqnarray}
For the quantity $\bm{B} =\sum_{k=0}^N \alpha_{k} \bm{s}_{k}$, our assumption (\ref{eq:condition_exsitence}) leads to
\begin{eqnarray}
 \bm{B} &=& \sum_{\sigma_j\neq \overline{\sigma}_j, \underline{\sigma}_j} \sum_{d=+,-} \left[ \bm{B} \right]_{\sigma_j, d} + \left[ \bm{B} \right]_{\overline{\sigma}_j, -} + \left[ \bm{B} \right]_{\underline{\sigma}_j, +}.
 \label{eq:decompose_B}
\end{eqnarray}
We also collectively write the Press-Dyson vectors of player $j$ by $\hat{\bm{T}}_j\left( \sigma_j \right) := \left( \hat{T}_j\left( \sigma_j | \bm{\sigma}^{\prime} \right) \right)_{\bm{\sigma}^\prime \in \mathcal{A}}$.
Below we construct ZD strategies for the case $M_j>2$ and the case $M_j=2$ separately.
(Because of the existence of two different actions $\overline{\sigma}_j, \underline{\sigma}_j$, $M_j$ must be greater than $1$.)
\begin{enumerate}[(i)]
\item $M_j>2$\\
For the case, we set a strategy of player $j$ as
\begin{eqnarray}
 \hat{\bm{T}}_j\left( \overline{\sigma}_j \right) &=& \frac{1}{W} \left( \sum_{\sigma_j \neq \overline{\sigma}_j, \underline{\sigma}_j} \left[ \bm{B} \right]_{\sigma_j, +} + \left[ \bm{B} \right]_{\overline{\sigma}_j, -} \right) \nonumber \\
 \hat{\bm{T}}_j\left( \underline{\sigma}_j \right) &=& - \frac{1}{W} \left( \sum_{\sigma_j \neq \overline{\sigma}_j, \underline{\sigma}_j} \left[ \bm{B} \right]_{\sigma_j, -} + \left[ \bm{B} \right]_{\underline{\sigma}_j, +} \right) \nonumber \\
 \hat{\bm{T}}_j\left( \sigma_j \right) &=& \frac{1}{W} \left( - \left[ \bm{B} \right]_{\sigma_j, +} + \left[ \bm{B} \right]_{\sigma_j, -} - \frac{1}{M_j-2} \left[ \bm{B} \right]_{\overline{\sigma}_j, -} \right. \nonumber \\
 && \qquad \left. + \frac{1}{M_j-2} \left[ \bm{B} \right]_{\underline{\sigma}_j, +}  \right) \quad \left( \sigma_j \neq \overline{\sigma}_j, \underline{\sigma}_j \right),
 \label{eq:strategy_Mg2}
\end{eqnarray}
where we have defined
\begin{eqnarray}
 W &:=& \max_{\bm{\sigma}\in \mathcal{A}} \left| B(\bm{\sigma}) \right| \neq 0.
 \label{eq:W}
\end{eqnarray}
We can easily check that these vectors indeed satisfy the condition of strategies (\ref{eq:property_strategy}) and (\ref{eq:condition_strategy}).
In addition, the condition (\ref{eq:PD_normalized}) is also satisfied because
\begin{eqnarray}
 \sum_{\sigma_j} \hat{\bm{T}}_j\left( \sigma_j \right) &=& \bm{0}.
\end{eqnarray}
Furthermore, the strategy (\ref{eq:strategy_Mg2}) satisfies
\begin{eqnarray}
  \hat{\bm{T}}_j\left( \overline{\sigma}_j \right) - \hat{\bm{T}}_j\left( \underline{\sigma}_j \right) &=& \frac{1}{W} \bm{B},
\end{eqnarray}
where we have used Eq. (\ref{eq:decompose_B}).
Therefore, the strategy (\ref{eq:strategy_Mg2}) is a ZD strategy.
\item $M_j=2$\\
For the case, we remark that the two actions of player $j$ are $\overline{\sigma}_j$ and $\underline{\sigma}_j$.
We set a strategy of player $j$ as
\begin{eqnarray}
 \hat{\bm{T}}_j\left( \overline{\sigma}_j \right) &=& \frac{1}{W} \left( \left[ \bm{B} \right]_{\overline{\sigma}_j, -} + \left[ \bm{B} \right]_{\underline{\sigma}_j, +} \right) \nonumber \\
  \hat{\bm{T}}_j\left( \underline{\sigma}_j \right) &=& - \hat{\bm{T}}_j\left( \overline{\sigma}_j \right),
 \label{eq:strategy_M2}
\end{eqnarray}
where $W$ is defined by Eq. (\ref{eq:W}).
We can easily check that these vectors indeed satisfy the condition of strategies (\ref{eq:PD_normalized}),  (\ref{eq:property_strategy}), (\ref{eq:condition_strategy}).
In addition, the strategy (\ref{eq:strategy_M2}) satisfies
\begin{eqnarray}
  \hat{\bm{T}}_j\left( \overline{\sigma}_j \right) &=& \frac{1}{W} \bm{B},
\end{eqnarray}
where we have used Eq. (\ref{eq:decompose_B}).
Therefore, the strategy (\ref{eq:strategy_M2}) is a ZD strategy.
\end{enumerate}
$\Box$

\subsection{Example}
\label{subsec:example}
In this subsection, we construct a ZD strategy for the case of the repeated prisoner's dilemma game.
The prisoner's dilemma game is a two-player two-action symmetric game with following payoffs:
\begin{eqnarray}
  \bm{s}_1 &:=&  \left( s_1(1, 1), s_1(1, 2), s_1(2, 1), s_1(2, 2) \right)^\mathsf{T} \nonumber \\
  &=& \left( R, S, T, P \right)^\mathsf{T} \nonumber \\
   \bm{s}_2 &:=&  \left( s_2(1, 1), s_2(1, 2), s_2(2, 1), s_2(2, 2) \right)^\mathsf{T} \nonumber \\
    &=& \left( R, T, S, P \right)^\mathsf{T},
\end{eqnarray}
where $T>R>P>S$ and $2R>T+S$.
(The actions $1$ and $2$ correspond to cooperation and defection, respectively.)
If we consider the quantity $\bm{B} =\sum_{k=0}^2 \alpha_{k} \bm{s}_{k}$ with $\alpha_1=0$ and $\alpha_2=1$, 
\begin{eqnarray}
  \bm{B} &=&  \left( R+\alpha_0, T+\alpha_0, S+\alpha_0, P+\alpha_0 \right)^\mathsf{T}.
\end{eqnarray}
Then, if we choose $\alpha_0$ as $\alpha_0 \in [-R, -P]$, we find that the actions $1$ and $2$ of player $1$ satisfy the condition of Theorem \ref{thr:existence} as $\overline{\sigma}_1=2$ and $\underline{\sigma}_1=1$.
Therefore, we conclude that the repeated prisoner's dilemma game contains at least one ZD strategy, which is a well-known result.
By using the construction method in the proof of Theorem \ref{thr:existence}, $\bm{B}$ is decomposed into
\begin{eqnarray}
  \left[ \bm{B} \right]_{2, -} &=&  \left( 0, 0, S+\alpha_0, P+\alpha_0 \right)^\mathsf{T} \nonumber \\
  \left[ \bm{B} \right]_{1, +} &=& \left( R+\alpha_0, T+\alpha_0, 0, 0 \right)^\mathsf{T},
\end{eqnarray}
and the ZD strategy is
\begin{eqnarray}
 \hat{\bm{T}}_1\left( 2 \right) &=& \frac{1}{W} \bm{B} \nonumber \\
  \hat{\bm{T}}_1\left( 1 \right) &=& - \hat{\bm{T}}_1\left( 2 \right)
\end{eqnarray}
with $W := \max_{\bm{\sigma}\in \mathcal{A}} \left| B(\bm{\sigma}) \right|$.
It should be noted that this ZD strategy is called the equalizer strategy and it unilaterally enforces $\left\langle s_{2} \right\rangle^{*}=-\alpha_0$ \cite{PreDys2012}.

\subsection{Relation to other class of stage games}
\label{subsec:class}
Theorem \ref{thr:existence} can be used to define a class of stage games where ZD strategies exist.
In this paper, we call this class \emph{ZD games}.
A natural question is the relation between ZD games and other classes of stage games, such as potential games and totally symmetric games.

Here, we restrict our attention to two-player symmetric games.
In other words, the payoffs satisfy $s_2(\sigma_1, \sigma_2)=s_1(\sigma_2, \sigma_1)$ $(\forall \bm{\sigma})$.
We also write the set of actions as $A_1=A_2=A:=\{ 1, \cdots, L \}$, where $L$ is the common number of actions.
We first introduce the concept of generalized rock-paper-scissors games.
\begin{definition}[\cite{DOS2012a,DOS2012b}]
\label{def:gRPS}
A stage game is \emph{a generalized rock-paper-scissors (gRPS) game} if it contains at least one subset of the action space $A^\prime \subseteq A$ such that for all $\sigma_1 \in A^\prime$ there exists $\sigma_2 \in A^\prime$ such that $s_1^{(\mathrm{A})}\left( \sigma_1, \sigma_2 \right) < 0$, where $s_1^{(\mathrm{A})} (\sigma_1, \sigma_2) := \left[ s_1 (\sigma_1, \sigma_2) - s_1 (\sigma_2, \sigma_1) \right]/2$ is an anti-symmetric part of $s_1$.
\end{definition}

We also call the complementary set of gRPS games in all two-player symmetric games as \emph{non-gRPS games}.
We now prove the following theorem on the relation between non-gRPS games and ZD games.
\begin{theorem}
\label{thr:ngRPS}
If a stage game is not a gRPS game, then it is a ZD game.
\end{theorem}

\textit{Proof.}
If a stage game is not a gRPS game, then, for all $A^\prime \subseteq A$, there exists an action $\sigma_1 \in A^\prime$ such that $s_1^{(\mathrm{A})}\left( \sigma_1, \sigma_2 \right) \geq 0$ $\left( \forall \sigma_2 \in A^\prime \right)$.
Such action $\sigma_1$ is an unbeatable action in $A^\prime$.
It should be noted that $A^\prime$ can be $A$.
We now construct a series of unbeatable actions as follows.
First, $\sigma^{*(1)}$ is an unbeatable action when the action space is $A$, that is, 
\begin{eqnarray}
  s_1^{(\mathrm{A})}\left( \sigma^{*(1)}, \sigma_2 \right) &\geq& 0 \quad \left( \forall \sigma_2 \in A \right).
\end{eqnarray}
Then, for $2\leq l \leq L$, we recursively define the set $A^{(l)}:=A\backslash \left\{ \sigma^{*(1)}, \cdots, \sigma^{*(l-1)} \right\}$ and an action $\sigma^{*(l)}\in A^{(l)}$ such that
\begin{eqnarray}
  s_1^{(\mathrm{A})}\left( \sigma^{*(l)}, \sigma_2 \right) \geq 0 \quad \left( \forall \sigma_2 \in A^{(l)} \right).
  \label{eq:sigma*}
\end{eqnarray}
We also write $A^{(1)}:=A$.
We remark that such series $\left\{ \sigma^{*(l)} \right\}_{l=1}^L$ is well-defined due to the property of non-gRPS games.

Because $s_1^{(\mathrm{A})} (\sigma_1, \sigma_2) = \left[ s_1 (\sigma_1, \sigma_2) - s_2 (\sigma_1, \sigma_2) \right]/2$ for two-player symmetric games, we can see that $\underline{\sigma}_1=\sigma^{*(1)}$ in the condition of Theorem \ref{thr:existence}:
\begin{eqnarray}
  s_1\left( \sigma^{*(1)}, \sigma_2 \right) - s_2\left( \sigma^{*(1)}, \sigma_2 \right) &\geq& 0 \quad \left( \forall \sigma_2 \in A \right).
  \label{eq:ngRPS_under}
\end{eqnarray}
Next, we prove that $\overline{\sigma}_1=\sigma^{*(L)}$:
\begin{eqnarray}
  s_1\left( \sigma^{*(L)}, \sigma_2 \right) - s_2\left( \sigma^{*(L)}, \sigma_2 \right) &\leq& 0 \quad \left( \forall \sigma_2 \in A \right).
  \label{eq:ngRPS_over}
\end{eqnarray}
Assume to the contrary that
\begin{eqnarray}
  s_1^{(\mathrm{A})}\left( \sigma^{*(L)}, \sigma_2 \right) &>& 0 \quad (\exists \sigma_2 \in A).
\end{eqnarray}
(Because $s_1^{(\mathrm{A})}$ is an anti-symmetric part, $\sigma_2\neq \sigma^{*(L)}$.)
This is rewritten as
\begin{eqnarray}
  s_1^{(\mathrm{A})}\left( \sigma_2, \sigma^{*(L)} \right) &<& 0 \quad (\exists \sigma_2 \in A).
\end{eqnarray}
However, since $\sigma_2 \in \left\{ \sigma^{*(1)}, \cdots, \sigma^{*(L-1)} \right\}$ and $\sigma^{*(L)}\in A^{(l)}$ for $1\leq l \leq L$, this contradicts with Eq. (\ref{eq:sigma*}).
Therefore, we conclude that Eq. (\ref{eq:ngRPS_over}) indeed holds.
We now find that Eqs. (\ref{eq:ngRPS_under}) and (\ref{eq:ngRPS_over}) correspond to the condition for the existence of ZD strategies in Theorem \ref{thr:existence}.
We remark that a linear relation enforced by the ZD strategy is $\left\langle s_{1} \right\rangle^{*}=\left\langle s_{2} \right\rangle^{*}$.
$\Box$

It should be noted that an unbeatable imitation strategy exists if and only if a stage game is not a gRPS game \cite{DOS2012b}.
Theorem \ref{thr:ngRPS} also constructs an unbeatable ZD strategy, which unilaterally enforces $\left\langle s_{1} \right\rangle^{*}=\left\langle s_{2} \right\rangle^{*}$, for non-gRPS games.
Both results imply that, in non-gRPS games, it is not easy for players to exploit the opponent.
We also note that two-player symmetric potential games are a subset of non-gRPS games \cite{DOS2012b}.
Therefore, our result directly leads to the existence of ZD strategies in two-player symmetric potential games \cite{Ued2022}.

We finally remark that the converse of Theorem \ref{thr:ngRPS} is not true.
That is, ZD strategies can exist for some gRPS games.
An example is a game in Table \ref{table:gRPS_ZD}, which is a modified version of the RPS game.
\begin{table}[tb]
\caption{A gRPS game with a ZD strategy.}
  \begin{tabular}{|c|ccccc|} \hline
    & $R$ & $P$ & $S$ & $\overline{\sigma}$ & $\underline{\sigma}$ \\ \hline
   $R$ & 0,0 & -1,1 & 1,-1 & 0,0 & 0,0 \\
   $P$ & 1,-1 & 0,0 & -1,1 & 0,0 & 0,0 \\
   $S$ & -1,1 & 1,-1 & 0,0 & 0,0 & 0,0 \\
   $\overline{\sigma}$ & 0,0 & 0,0 & 0,0 & 0,0 & 0,0 \\
   $\underline{\sigma}$ & 0,0 & 0,0 & 0,0 & 0,0 & 0,0 \\ \hline
  \end{tabular}
  \label{table:gRPS_ZD}
\end{table}
Although this game contains a gRPS cycle when $A^\prime = \{ R, P, S \}$, this game is also a ZD game, since $\overline{\sigma}$ and $\underline{\sigma}$ are regarded as the two actions in Theorem \ref{thr:existence}.

\section{Concluding Remarks}
\label{sec:conclusion}
In this paper, we have provided the necessary and sufficient condition for the existence of ZD strategies in repeated games (Theorem \ref{thr:existence}).
This condition exactly means the existence of two actions which unilaterally increases and decreases the total value of $\sum_{k=0}^N \alpha_{k} s_{k}$, respectively.
We have now found that such property is necessary for unilateral control of payoffs by ZD strategies.
In fact, we can easily check that the rock-paper-scissors game does not contain the two actions as in Theorem \ref{thr:existence}, which leads to the absence of ZD strategies \cite{UedTan2020}.
From another point of view, stage games satisfying the condition of Theorem \ref{thr:existence} can be regarded as a class allowing the existence of ZD strategies.
We also provided the relation between this class of stage games (ZD games) and non-gRPS games for the case of two-player symmetric games.
Further investigation on the relation between ZD games and other classes of stage games is needed.

We have investigated only the situation that a discounting factor is $\delta=1$ and monitoring is perfect.
In general, the set of possible ZD strategies decreases as $\delta$ decreases and monitoring becomes imperfect \cite{HRZ2015,IchMas2018,UedTan2020,MMI2021}.
Particularly, the existence of ZD strategies in games with imperfect monitoring will be highly dependent on the set of signals of each player.
Investigation of the necessary and sufficient condition for the existence of ZD strategies in games with discounting and imperfect monitoring is an important subject of future work.
It would be interesting if our result can be applied for the existence of memory-$m$ ZD strategies with $m\geq 2$ \cite{Ued2021b}.

\begin{acknowledgment}
This study was supported by JSPS KAKENHI Grant Number JP20K19884 and Inamori Research Grants.
\end{acknowledgment}


\bibliographystyle{jpsj}
\bibliography{conditionZDS}

\end{document}